\documentclass{article}%iopart}
\usepackage{graphicx}
\usepackage{amsmath}
\begin{document}
\title{Melting of hexagonal Skyrmion states in chiral magnets}
\maketitle
\author{M C Ambrose \footnote{1. School of Physics, The University of Western Australia, 35 Stirling Hwy, Crawley 6009, Australia \\
ambrom01@student.uwa.edu.au} , R L Stamps \footnote{ 2. SUPA School of Physics and Astronomy, University of Glasgow, Glasgow G12 8QQ, United Kingdom \\
Robert.Stamps@glasgow.ac.uk} }
\begin{abstract}
Skyrmions are spiral structures observed in thin films of certain  magnetic materials \cite{Uchida20012006}. Of the phases allowed by the crystalline symmetries of these materials \cite{PhysRevB.80.054416} only the hexagonally packed phases ($SC_h$) has been observed. Here the melting of the $SC_h$ phase is investigated using Monte Carlo simulations. In addition to the usual measure of Skyrmion density chiral charge, a morphological measure is considered. In doing so it is shown that the low temperature reduction in chiral charge is associated with a change in Skyrmion profiles rather than Skyrmion destruction. At higher temperatures the loss of six fold symmetry is associated with the appearance of elongated Skyrmions that disrupt the hexagonal packing. 
\\
Pacs :12.39.Dc, 75.70.-i
\end{abstract}

%\submitto{\NJP}

\newpage
\section{Introduction}
\begin{figure}[htb] 
\centering
\includegraphics[scale = .075]{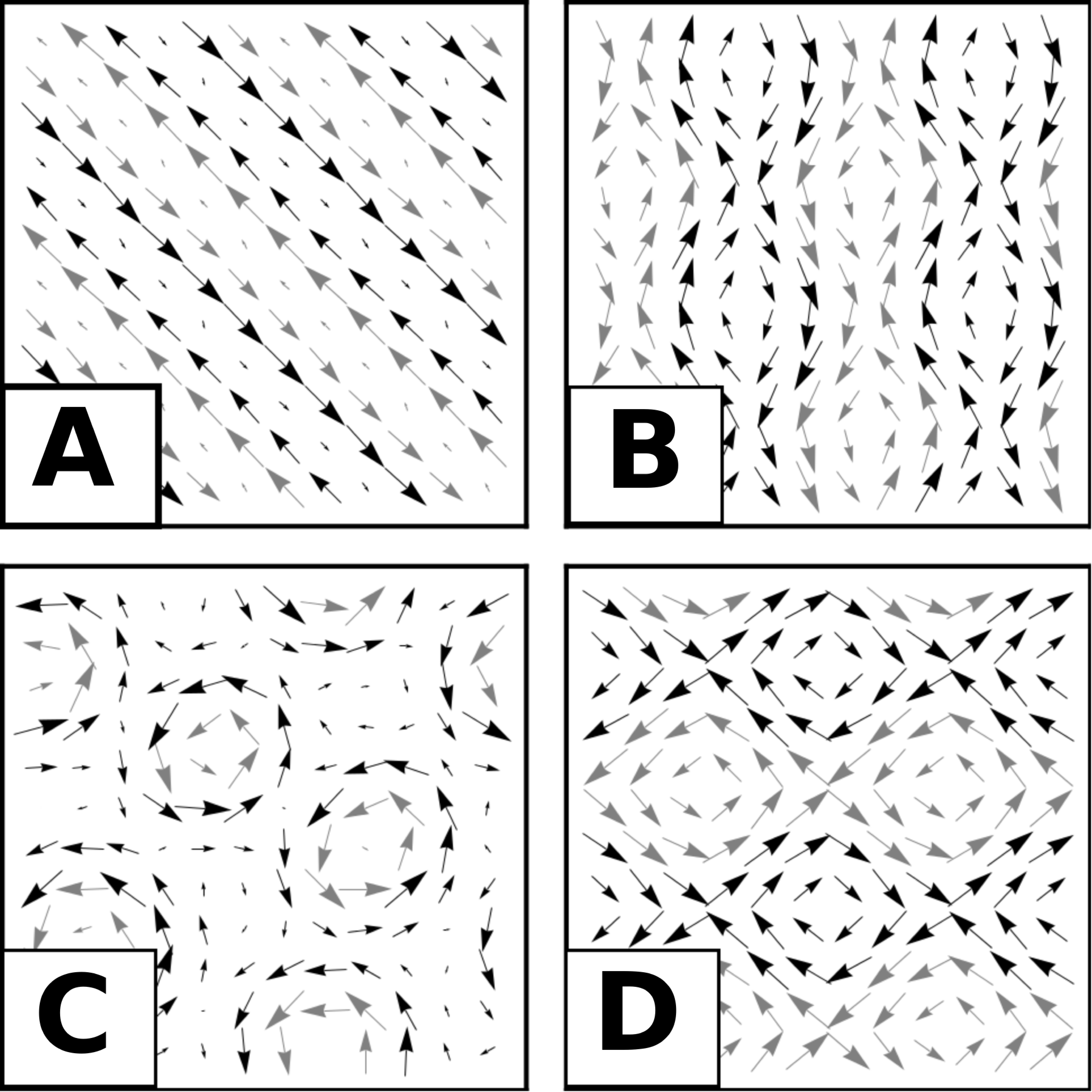}
   \caption{Projection of spins into the x-y plane, black arrows indicate that $s_i^{\: z} > 0 $ while gray arrows indicate $s_i^{\: z} < 0 $  A: Helix State; B: $SC_1$ Skyrmion; C: $SC_2$ Skyrmion; D: $SC_h$ Skyrmion. }
\label{states}
\end{figure}

Following early work by Pauli a variety of non-linear sigma models were suggested as models for baryons \cite{Pauli,Skyrme16091958,Skyrme08091959}. One of the first soluble models was found by Skyrme who considered rotational variables \cite{Skyrme07021961} and the resultant unit field spiral solitons are referred to as Skyrmions. Two dimensional Skyrmions appeared as the fundamental excitation from a spin polarized two dimensional electron gas. It was suggested that defects could localize Skyrmions giving rise to the fractional quantum hall effect \cite{PhysRevLett.62.82, PhysRevLett.64.1313,PhysRevB.47.16419}. Observations in GaAs quantum well systems using NMR \cite{PhysRevLett.74.5112} and magnetoabsorption spectroscopy \cite{PhysRevLett.76.680} confirmed the presence of quasi-particles with charges consistent with theoretical predictions. More recently measurements have been made in GaAs using NMR relaxation \cite{PhysRevLett.94.196803}, spin wave absorption \cite{PhysRevLett.100.086806} and microwave absorption \cite{PhysRevLett.104.226801}, that are consistent with the presence of a predicted Skyrmion lattice \cite{PhysRevLett.75.2562}. Skyrmions were first proposed as a possible magnetic spin texture by Bogdanov and Hubert who showed that Dzyaloshinskii-Morya exchange terms could lead to stabilization of Skyrmion crystals in chiral magnets \cite{Bogdanov1994255}. Recently the real space measurements of the two dimension analogue these spiral structures has been made at low temperature in thin films of $\text{Fe}_{0.5}\text{Co}_{0.5}\text{Si}$ \cite{Uchida20012006,RSSkirmions} and close to room temperature in FeGe\cite{HiTSkyrme}, along with the measurement of spin torque effects at very low current densities \cite{Jonietz17122010}, these measurements have ignited interest in Skyrmions as a possible candidate for magnetic storage\cite{Storage}. \\
In two dimensions Skyrmions exist as a vortex structure modulated by a changing perpendicular component. Consider a field of unit length $\hat{s}(\vec{x})$ with $\vec{x}$ two dimensional and $\hat{s}(\vec{x})$ taking the usual polar representation \footnote{The zenith measured with respect to the $z$ direction perpendicular to the plane}. Away from an isolated Skyrmion the field is described by zenith angle $\theta = \pi$. Taking radial coordinates $(\rho,\psi)$ for $\vec{x}$ a Skyrmion can be described as a vortex in the azimuthal angle $\phi(\rho,\psi)= \psi-\pi/2$ modulated by a radial varying zenith angle $\theta(\rho,\psi)= \theta(\rho)$ such that $\theta(0)=0$ and $\theta(R)=\pi$. An example of close packed Skyrmions projected into the plane is shown in figure \ref{states} (c) where gray(black) arrows indicate spins with perpendicular component are pointing up(down). In order for such an excitation to be stable one requires a Hamiltonian with either fourth order derivatives of the field \cite{Skyrme07021961} or terms lacking inversion symmetry\cite{Bogdanov1994255}. Lack of inversion symmetry can be found in a variety of crystal structures, in particular B20 crystal compounds with crystallographic point group $23$ \footnote{in Hermann Mauguin notation}, in which Dzyaloshinskii-Morya (DM) coupling is present. Yi et a.l \cite{PhysRevB.80.054416} calculated the $T=0$ phase diagram as a function of external magnetic field and anisotropy for such a two dimensional magnet using Monte-Carlo simulation. Five different states were identified: The high field saturated ferromagnetic phase, a helical state (figure \ref{states}(a)) and three chiral states shown. The chiral crystal states were labeled according to their rotational symmetries: $SC_1$ (figure \ref{states}(b)) has a two fold rotational symmetry, $SC_2$ (figure \ref{states}(d)) shows a four fold symmetry and $SC_h$ (figure \ref{states} (c)) a six fold symmetry \footnote{This notation should not be confused with the notation some authors in which Skyrmion states are labeled as either SkX or SkG refering to whether the $SC_h$ has formed a close packed crystal (SkX) or are freely moving (SkG)}. In what follows we restrict our attention the the $SC_h$ type skyrmions.\\
Analytically the ground state properties of this two dimensional case have been studied using a Landau-Ginzburg formalism \cite{PhysRevB.82.094429}. The nature of the phase transition has already been investigated using combinatorics \cite{PhysRevB.83.100408}. Here the authors considered the striped helical state structure shown in figure \ref{states} (a) as the ground state. In the presence of an external field perpendicular to the plane of the sample, one direction of perpendicular magnetization is favored.  Since the stripe period is fixed by the ratio of  Dzyaloshinskii-Morya and ferromagnetic exchange coupling, spins reverse inside stripes anti-aligned with the field, breaking remaining anti-aligned areas into finite length stripes. Unlike the one dimensional periodicity of the ground state (giving zero chiral density) the ends of the terminating stripes form a 'meron'; a chiral half spiral. The Skyrmion crystal state $SC_h$ show in figure \ref{states}(c) is the state with maximum possible meron density. By considering the merons as particles with chiral charge, a free energy can be written and a qualitatively correct B-T phase diagram produced. While the assumptions of this model are reasonable low temperature, the possibility of Skyrmion profiles changing with temperature cannot be described. One expects, that as temperature is increased, thermal fluctuations will dominate and the particle description will no longer be valid. \\
Current Monte Carlo simulations have focused on obtaining phase diagrams, with phases identified according to symmetry properties and chiral density \cite{RSSkirmions, PhysRevB.80.054416}. As a system with textured phase, close packed Skyrmions represent three types of order: a net chiral charge, a six fold symmetry and a net magnetization. Here we use Monte Carlo simulations to investigate the melting of the $SC_h$ phase, focusing on the mechanism through which these various orders are destroyed. 
\section{Theory and Method}
Bak and Jensen proposed a continuum Hamiltonian \cite{HelixMagnets} to explain observations of helical structure in MnSi \cite{Ishikawa1977401,Ishikawa1976525,Kusaka1976925} and FeGe \cite{1402-4896-1-1-012}.
The model was adapted to a discrete lattice by Yi et al.  \cite{PhysRevB.80.054416}  and we begin with their Hamiltonian. In two dimensions and in our notation, it reads
\begin{equation} 
\label{ham}
\begin{split}
E = & -J/2 \sum_{\langle i,j \rangle} \vec{s}_i \cdot \vec{s}_j 
-  K/2 \sum_{i}  ((\vec{s}_i \times( \vec{s}_{i+a \hat{x}} -\vec{s}_{i-a\hat{x}})) \cdot \hat{x} 
+(\vec{s}_i \times( \vec{s}_{i+a\hat{y}} 
-   \vec{s}_{i-a\hat{y}}) ) \cdot \hat{y} ) 
\\
&+\sum_{i} \vec{H}\cdot\vec{s}_i
 + A_1 \sum_{i} \left((\vec{s}_i^{\: x})^4+(\vec{s}_i^{\: y})^4+(\vec{s}_i^{\: z})^4 \right) 
+A_2 \sum_{i}  \left( s_i^{\: x} s_{i+a\hat{x}}^{\: x} +  s_i^{\: y} s_{i+a\hat{y}}^{\: y} \right)
\end{split}
\end{equation} 
 For brevity we use single subscripts to represent  locations on a two dimensional square lattice with lattice constant $a$. The $\vec{s}_i$ are dimensionless spins at the vertices of this lattice representing the average magnetization of a small volume $V_i = a \times a \times t$ centered on vertex $i$, with $t$ this thickness of the film. $\vec{s}_{i+a\hat{x}}$ and $\vec{s}_{i-a\hat{x}}$ represent the closest spins to $\vec{s}_i$ in the $x$ direction (with analogous notation in the $y$ direction) and $\sum_{\langle i,j \rangle}$ indicates a sum over all pairs of nearest neighbors.
 $K$ is the strength of the Dzyaloshinskii-Morya 
exchange coupling, the sign of which determines the handedness of the Skyrmions; $J$ is the strength of the isotropic exchange coupling; $H$ is the strength of the applied magnetic field and $A_1$ and $A_2$ are anisotropies.
The ratio of exchange to Dzyaloshinskii-Morya strength determines the periodicity of Skyrmion lattice as $P= d a$ where $d$  is \cite{PhysRevB.80.054416, RSSkirmions}: 
\begin{equation}
\label{size}
d = 2  \pi \left(\arctan\left(\frac{2\sqrt{2}K}{4 J+ A_2}\right ) \right)^{-1}.
\end{equation}
\begin{table}[htb]
\centering
\begin{tabular}{ |l | c | }
    \hline
    Parameter & Value  \\ \hline \hline
     $ K/J$& 2.45  \\
    $A_1/J$& 0.5\\
    $A_2/J$& 0.5\\
    \hline
  \end{tabular}
\caption{Parameters used in the Monte Carlo simulation.}
\label{d6p}
\end{table}
Since the computation time to complete a Monte Carlo step scales with the number of spins, simulating large systems is typically associated with long computation times. In order to decrease this computation time GPU parallel programming is employed. The Hamiltonian includes only short range interactions so a parallel checkerboard type algorithm similar to those described by Weigel and Yavorskii \cite{Weigel201192,Weigel20111833} is used.
\subsection{Ground State Packing}
\begin{figure}[htb]
\centering
\includegraphics[scale = .25]{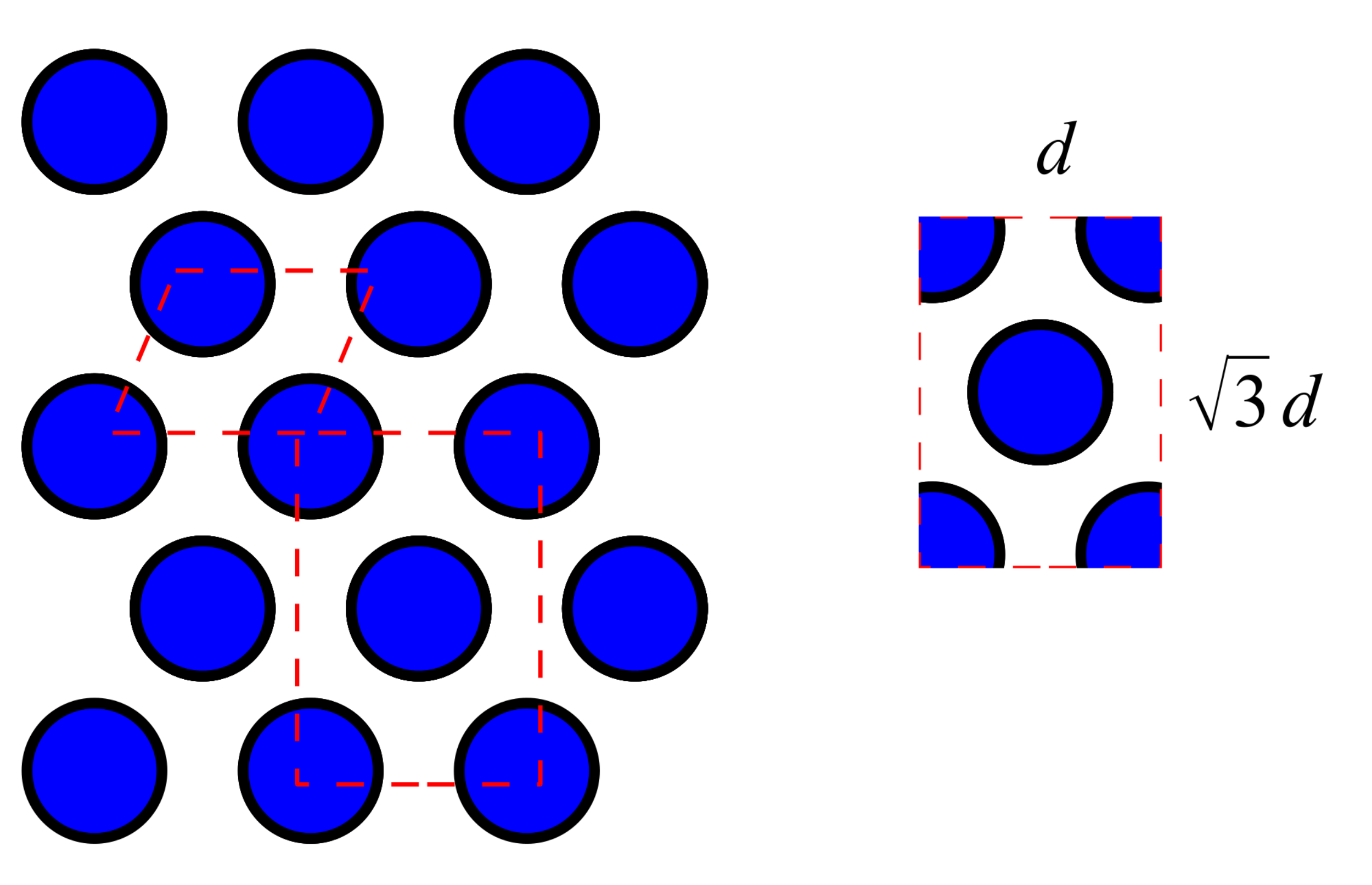}
\caption{Left: Unit cells that span a plane with hexagonal symmetry; Right: the dimensions of a rectangular unit cell that can be used to tile a finite system with periodic boundary conditions. }
\label{Stack}
\end{figure}

\begin{figure}[htb]
\centering
\includegraphics[scale = .25]{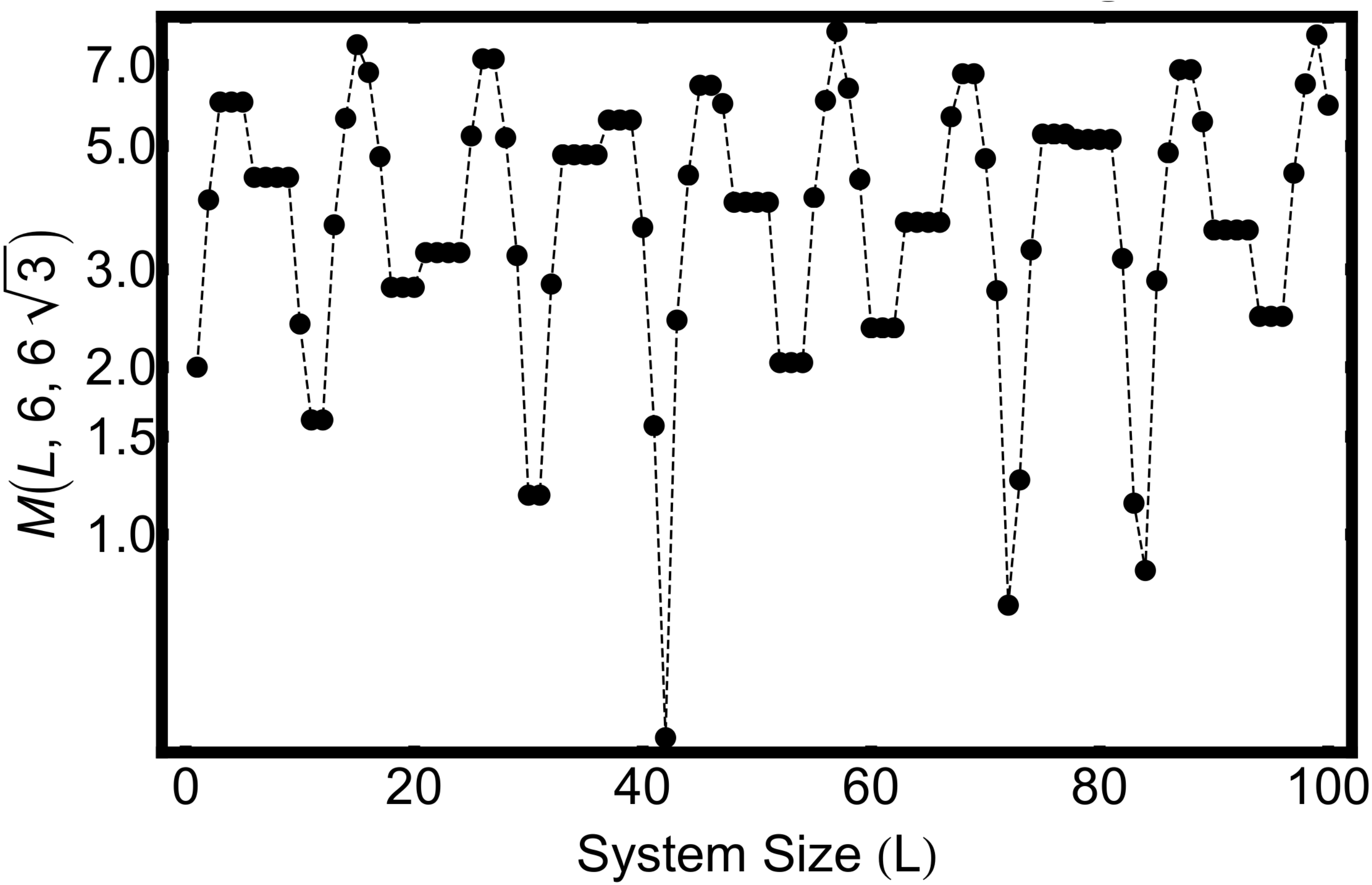}
\caption{ Log plot of $M(L,6,6 \sqrt{3})$ showing the minimum at $L=42$. This minimum corresponds corresponds to the tiling that best minimizes finite size effects introduced by periodic boundary conditions. }
\label{Min}
\end{figure}

In the interest of comparison with previous results \cite{RSSkirmions, PhysRevB.80.054416} we choose the material parameters so that the expected dominant wave length will be $d=6$. Parameters used are given in table \ref{d6p}, where dimensionless quantities are given by normalizing against $J$. In what follow temperature and energy will be given as  $\mathcal{T}=(k_B T)/J$  and energy as $\mathcal{E}=E/ J$. The magnetic field is fixed at $  H/J = 1.875 $ to ensure that at $\mathcal{T}=0$ the system forms close packed Skyrmions. \\
Some care should be taken when choosing the finite dimension ($L$) of the spin lattice for these calculations. 
For Skyrmions the ground state is periodic, consisting of hexagonally close packed Skyrmions, hence the repeated unit cell of the crystal lattice is not square. In order to tile a pattern with six fold symmetry one usually selects a unit cell that is a parallelogram defined by the vectors $v_1= (d,0)$ and $v_2  =  (d/2, \sqrt{3}d/2)$, where $d$ is the spacing between six fold symmetry axes (in this case Skyrmion cores). The resulting parallelogram is indicated in figure \ref{Stack}.
This parallelogram unit cell is space filling and  can be used to cover an infinite plane.
\\ Our simulation uses a square lattice spanning a finite region with periodic boundary conditions. In order to tile a space with periodicity in the $x$ and $y$ directions we select a rectangle to act as the unit cell, also shown in figure \ref{Stack}. This unit cell is then a rectangle with width $d_1=d$ and height $d_2 = \sqrt{3}d$. The above parameters for which $d = 6$ correspond to a unit cell of dimensions $6 \times 10.3923$. If one picks a system size with periodic boundary conditions and length as multiples of $d$, the resultant state will be influenced in one direction by the forced periodicity. When deciding the system size one should attempt to find a system size that is a multiple of $d_1$ and $d_2$. With $d_2$ irrational this is not possible. Instead we consider the two lengths that define the unit cell $d_1 = 6$ and $d_2 =  6\sqrt{3}$. In order to measure the amount of mismatch between the system size $L$ and the unit cell we define a function:
\begin{equation}
\label{FracA}
\begin{split}
M(L,d_1,d_2) = &\text{min}(L-d_2\lfloor L/d_2 \rfloor,d_2-L+d_2\lfloor L/d_2 \rfloor)
\\
+&\text{min}(L-d_1\lfloor L/d_1 \rfloor,d_1-L+d_1\lfloor L/d_1 \rfloor)
\end{split}
\end{equation}
where each term in equation (\ref{FracA}) is a measure of how far L is from an exact multiple of $d_1$ or $d_2$. In the case of the hexagonal Skyrmion lattice where $d_2$ is irrational and $L$ is restricted to be a integer, the minima are not known a priori. In order to select a reasonable size, $M(L,6,6 \sqrt{3})$ was calculated for $L \in [1,100]$. As shown in figure \ref{Min}, the minimum value is found to be $L=42$.
\subsection{Analysis}
The close packed Skyrmion ground state consists of two types of order, a chiral charge associated with the creation of spiral structures and a long range order associated with the hexagonal packing. In addition to the Skyrmion crystal order the system has a net magnetization in response to the applied magnetic field.\\
Two techniques are employed in order to measure the number of Skyrmions in a state: The first is to consider the topological charge $Q$ of a Skyrmion , which is the sum of the local chiral charge over the area of a single Skyrmion \cite{PhysRevB.83.100408}. The charge density is:
\begin{equation}
\label{Chiral}
\rho_{i} = \frac{1}{4 \pi } \vec{s}_{i} \cdot \left ( (\vec{s}_{i+a\hat{x}} -\vec{s}_{i-a\hat{x}}) \times  (\vec{s}_{i+a\hat{y}} -\vec{s}_{i-a\hat{y}})  
\right )
\end{equation}
In order to define the charge $Q$ of a Skyrmion, one calculates the total charge density of the ground state and divides by the number of Skyrmions. In order to calculate the Skyrmion number at some finite temperature one divides the total charge by $Q$. \\
This method of counting Skyrmions assumes that the Skyrmion charge remains constant (i.e that the Skyrmion profile is not temperature dependent). For comparison, we define a second measure of Skyrmion number. Since the core of a Skyrmion points against the applied magnetic field one can binarize a state by applying $\Theta(1/2(\vec{s}_{i}.\hat{z} +1) -T_r)  $ ($\Theta$ is the Heaviside Theta function), where $T_r$ is some threshold between zero and one. In doing so one identifies spins with $\vec{s}_{i}.\hat{z} $ anti-parallel to the magnetic field as Skyrmion cores. One can then calculate the connected components of the resultant state. To do this the eight neighbors and next nearest neighbors to a given spin are considered connected if they are equal. Counting the total number of connected components gives a measure of the Skyrmion number. By comparing these two measures of Skyrmion number one can distinguish between reduction of topological charge due to Skyrmion destruction and alteration of Skyrmion profile.\\
In order to examine the long range order of a state the Fourier transform $\vec{S}_k = \sum_j \vec{s}_{j} \exp (\imath k \cdot j)$ is taken. When the the system is close packed $|\vec{S}_k|$ will have six satellite peaks corresponding to the hexagonal symmetry of the ground state.
\section{Results}
\begin{figure}[h!]
\centering
\includegraphics[scale = .25]{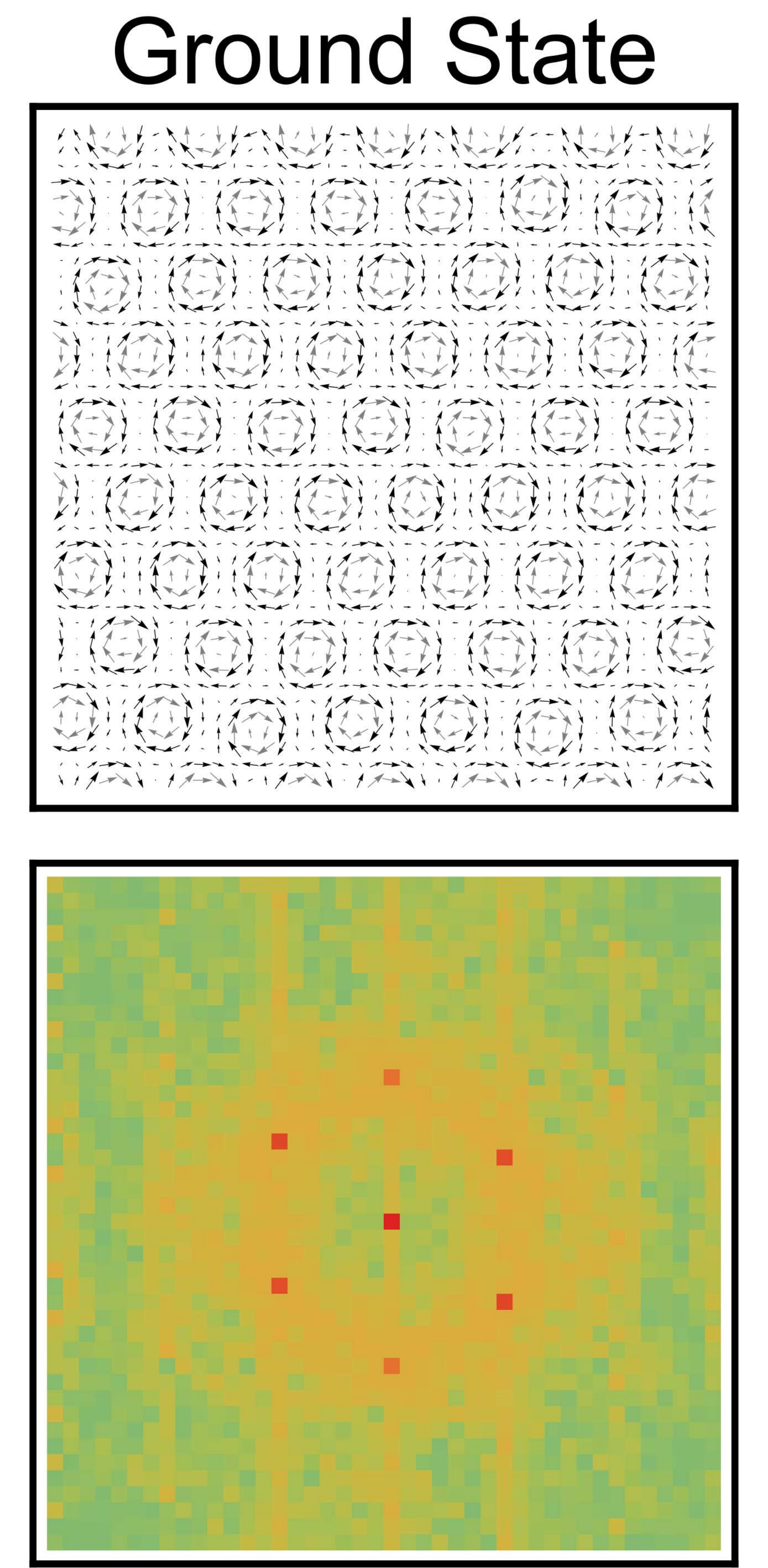}
\caption{The ground state spin configuration (top) and the intensity of the Fourier transform (bottom) showing peaks characteristic of the six fold translational symmetry of the hexagonal lattice.}
\label{GS}
\end{figure}

\begin{figure}[h!]
\centering
\includegraphics[scale = .25]{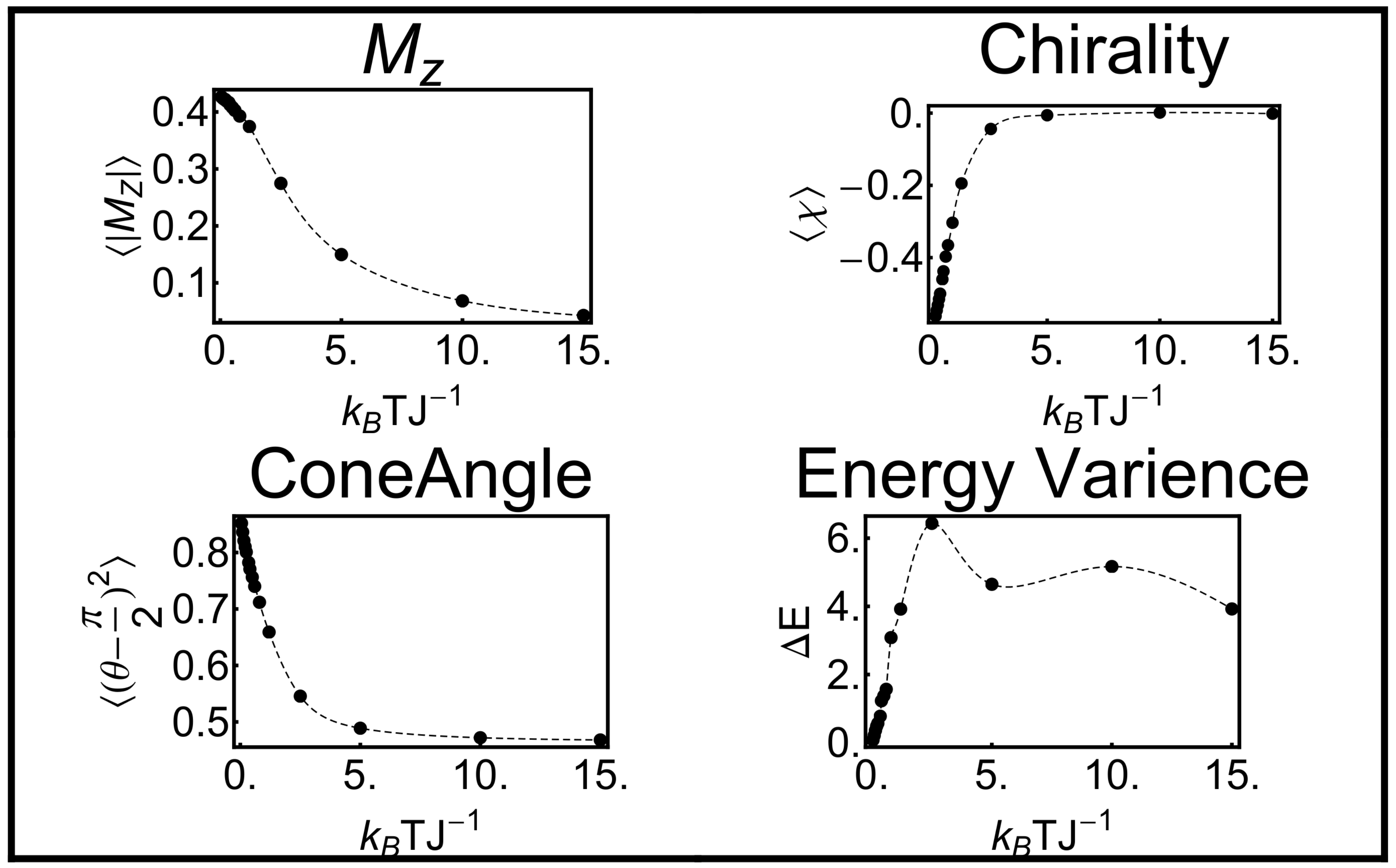}
\caption{Clockwise from top left: The perpendicular magnetization, the chiral charge, the cone angle and the energy variance.}
\label{Melt1}
\end{figure}

For each system size to be investigated the ground state is first calculated by starting the system in a random configuration and then reducing the temperature to zero over $10^7$ MC steps. The ground state for $L=42$ in shown in figure \ref{GS} along with the intensity profile of the Fourier transform $|\vec{S}_k | $. By selecting the system size $L$ to match the dimensions of the rectangular unit cell described above, it is possible to fit  an integer number of unit cells into the square array. 
For finite temperature results the system is evaluated at a constant temperature starting from the ground state with the first $10^7$ MC steps disregarded to allow the system to equilibrate. An ensemble of $100$ states is calculated and $280$ MC steps are taken between subsequent states to ensure that each state is selected independently. In figure \ref{Melt1} the order parameters of the system are shown indicating the low temperature transition into the Skyrmion state. The persistence of the magnetic response at temperatures greater than those required for the destruction of chiral order is seen through the magnetization $M_z$ and the cone angle. The loss of chiral structure occurs around $\mathcal{T}=3$, but the elevated cone angle and magnetic order persist at approximately $\mathcal{T}=8$.
The energy variance is double peaked indicating two broad transitions. At high temperatures the entropy dominates and there is no order. Below $\mathcal{T}=10$ there exists a region in which the cone angle increases from its high temperature value giving rise to a non zero magnetization. The external field ensures that as the temperature is decreased the magnetization increases. Below approximately $\mathcal{T}=2.5$ there is a transition into the Skyrmion state characterized by the sharp increase in the strength chiral density (negative values are indicative of left handed structure). Since the creation of Skyrmions prevents the system reaching saturation we note that the magnetization remains well below the saturation value.
\subsection{Loss of six fold order}
\begin{figure}[h!]
\centering
\includegraphics[scale = .35]{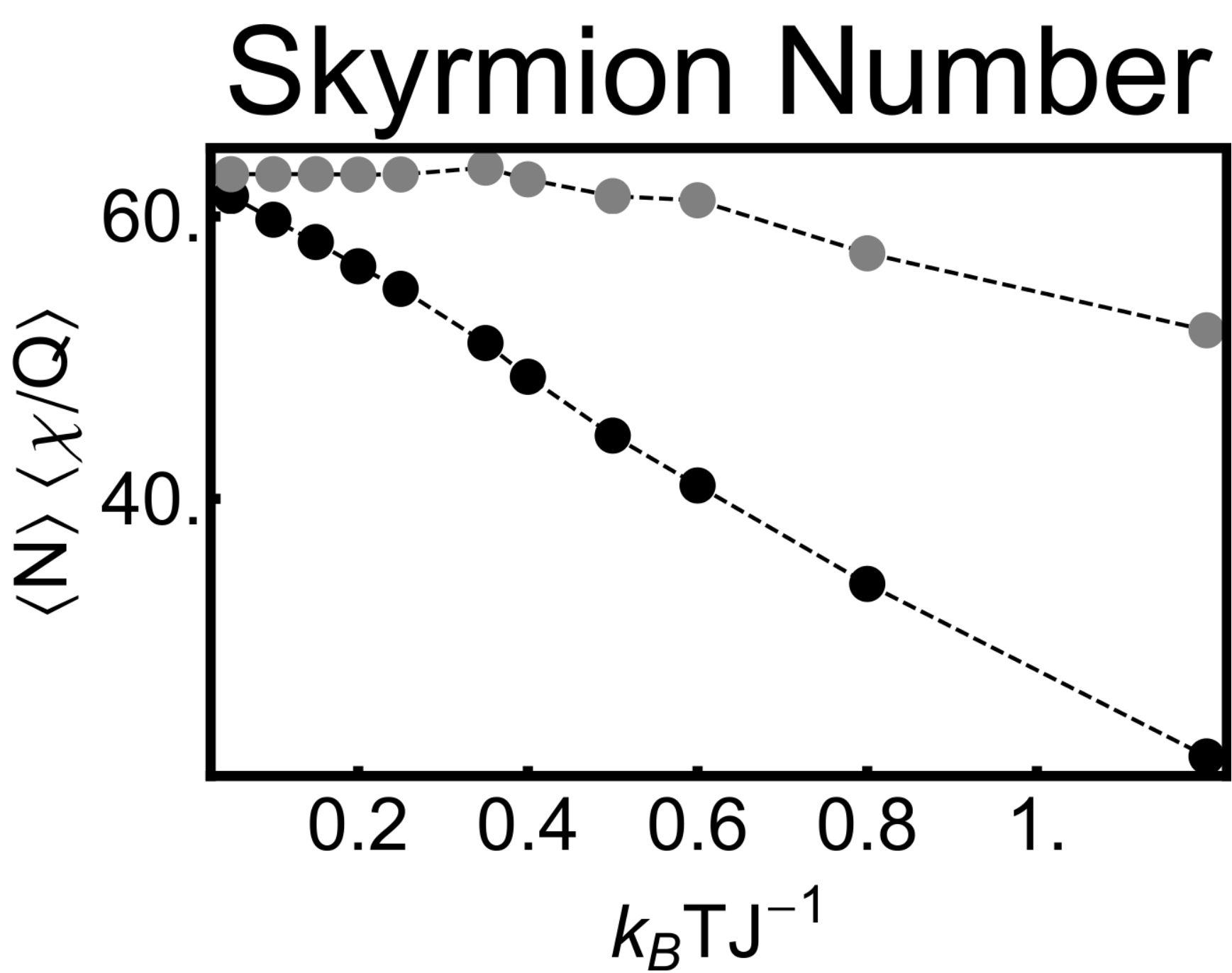}
\caption{Two measures of Skyrmion packing as a function of temperature: the connected reversed components (gray circles) and the ratio of total chiral charge to the charge of one Skyrmion in the ground state $\rho /Q$ (black circles). }
\label{Melt2}
\end{figure}

\begin{figure}[]
\centering
\includegraphics[scale = .15]{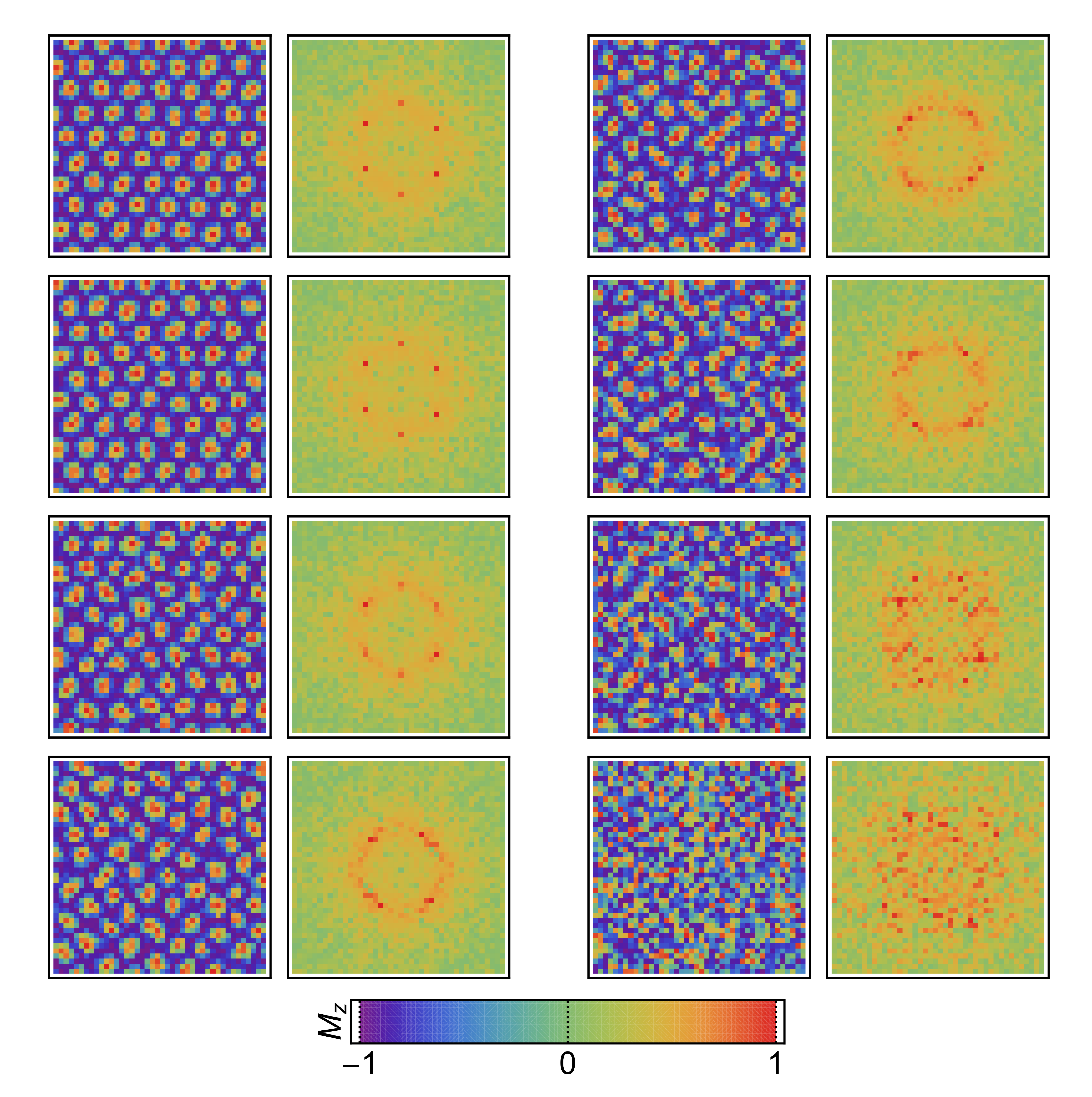}
\caption{The perpendicular components of $\vec{s}_{i}$ and the Fourier intensity $|\vec{S}_k|$ near the phase transition. Since the central ($k=0$) Fourier peak will significantly larger than the satellite peaks at finite temperature,  it is removed to give contrast. The Fourier plots then have their colors scaled with green representing zeros and and red representing the maximum value. Perpendicular components of $\vec{s}_{i}$ are colored according to the legend above. Left column from top: $\mathcal{T}=0.1$, $\mathcal{T}=0.2$, $\mathcal{T}=0.35$, $\mathcal{T}=0.4$  Right column from top: $\mathcal{T}=0.5$, $\mathcal{T}=0.8$, $\mathcal{T}=1.2$, $\mathcal{T}=2.5$. }
\label{Melt3}
\end{figure}

\begin{figure}[h!]
\centering
\includegraphics[scale = .5]{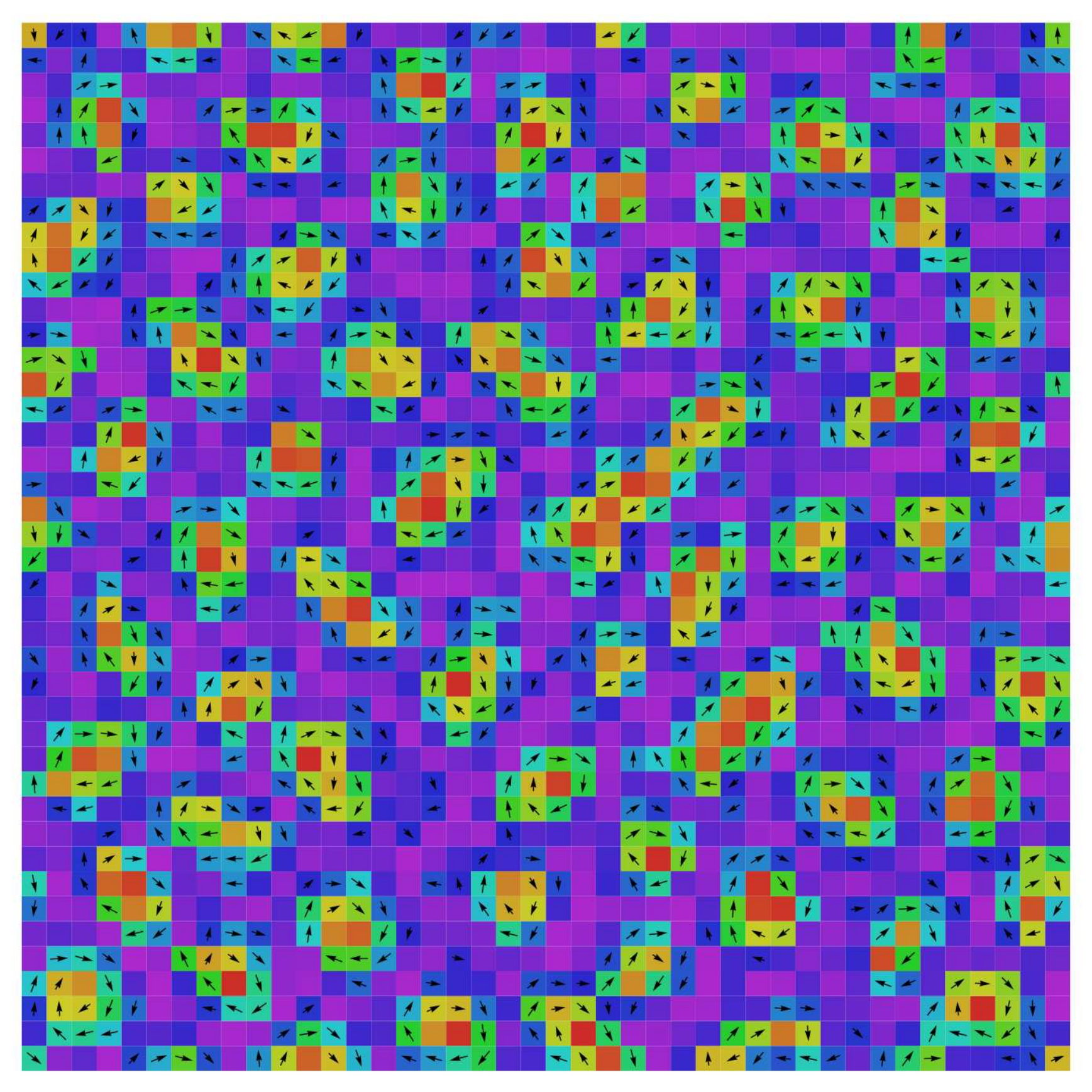}
\caption{An example of the elongated structures associated with the destruction of long range order at $\mathcal{T}=0.5$. Here the perpendicular components of the spins are represented by color as in figure \ref{Melt3}, in addition for spins parallel to the plane their direction is indicated with arrows.}
\label{Nice}
\end{figure}

\begin{figure}[h!]
\centering
\includegraphics[scale = .5]{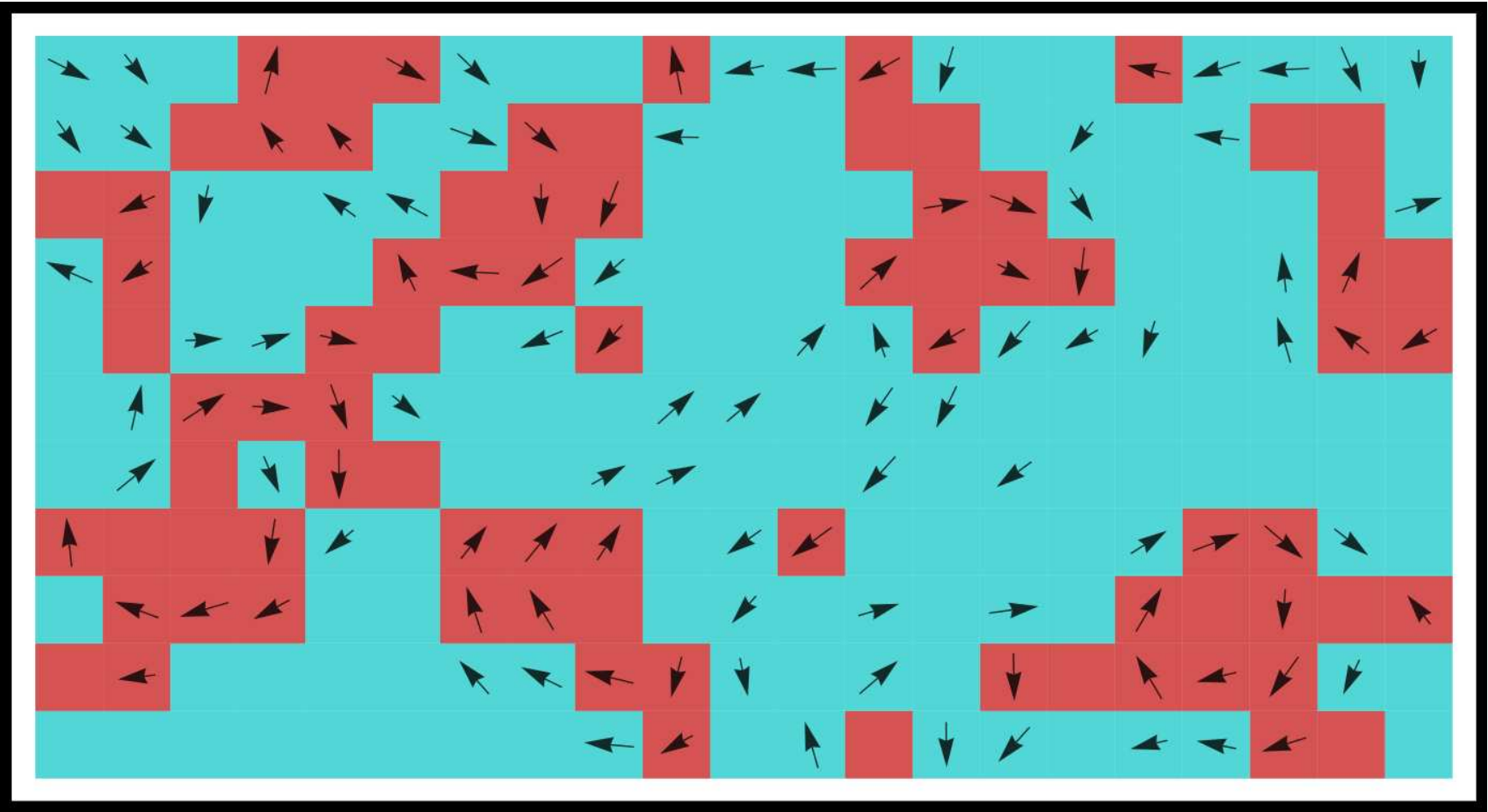}
\caption{An example of the elongated structures associated with the destruction of long range order at $\mathcal{T}=0.5$. Spins parallel to the plane have their direction indicated with arrows. Areas of high chiral density are indicated in red.  }
\label{Nice2}
\end{figure}

\begin{figure}[]
\centering
\includegraphics[scale = .1]{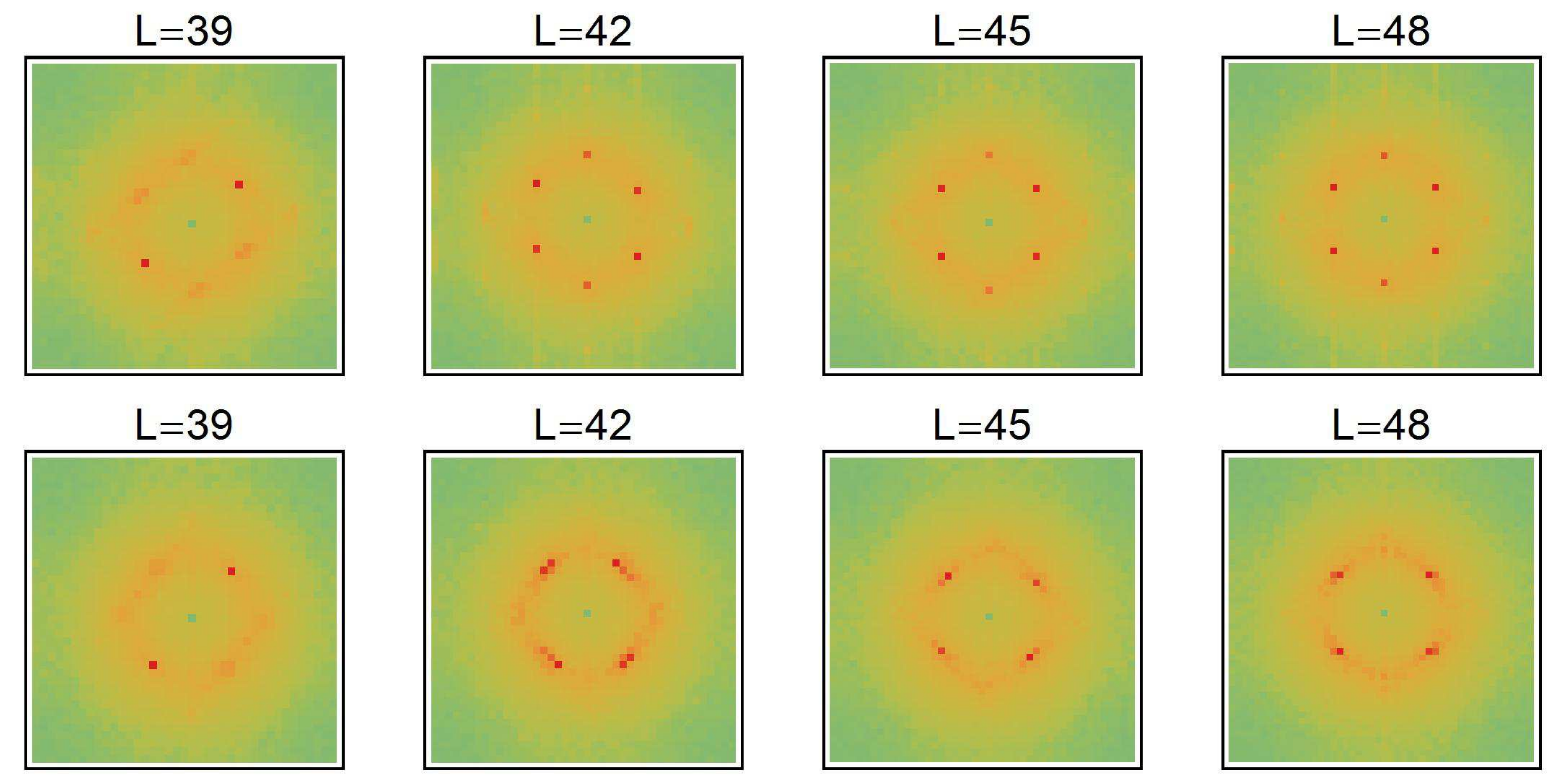}
\caption{ $\langle | \vec{S}_k | \rangle$ as function of size for $\mathcal{T}=0.2$(Top) and $\mathcal{T}=0.4$(Bottom). The plots are colored as in figure \ref{Melt3}. }
\label{SizeCom}
\end{figure}

In order to examine the loss of long range order we focus on the low temperature regime. In figure \ref{Melt2} the Skyrmion number is calculated using the total chiral charge and the method of counting cores described previously. While the chiral measure is reduced at all finite temperatures the number of reversed regions remains constant below about $\mathcal{T}=0.3$. This suggests that the initial loss of chiral charge is due to thermal distortion of the Skyrmion profiles rather than destruction of the Skyrmions themselves. 
In figure \ref{Melt3} an example state from ensembles at different temperatures is show along with $|\vec{S}_k|$, between $\mathcal{T}=0.1$ and $\mathcal{T}=2.5$. At $\mathcal{T}=0.4$ the strong six fold symmetry is compromised with peaks beginning to smear together. At $\mathcal{T}=0.5$ the peaks have smeared into a circle, indicating that a while a dominant length scale still exists sixfold translational symmetry is lost. At this temperature one can simultaneously observe the appearance of elongated regions of reversed magnetization. As temperature is further increased any order in the perpendicular components of $\vec{s}_{i}$ is destroyed. An enlarged example of a state at $\mathcal{T}=0.5$ is shown in figure \ref{Nice}. In addition to the perpendicular magnetic order, in plane magnetic order is shown as arrows. In addition to disrupting the close packing of Skyrmions these extended structures also reduce the chiral density described in equation \ref{Chiral}. To illustrate this, in figure \ref{Nice2} the longest of extended regions shown in figure \ref{Nice} is replotted with  information about perpendicular order omitted. Color is used to indicate the areas of highest chiral density, blue represents areas of low density and and red indicates areas of high density. Here only the ends of the structure contribute significantly.
\\The presence of the low temperature (below $ \mathcal{T}=0.3$) distortion of Skyrmion profile and the appearance elongated regions might indicate a change of the dominant length scale. While the difficulty associated with finite size effects was minimized by considering the dimensions of the ground state unit cells, this doesn't necessarily ensure that the effects are negligible at finite temperatures. Several other comparable system sizes were investigated for $\mathcal{T}=0.2$ and $\mathcal{T}=0.4$. The resulting energies are given in table \ref{SizeEng}, where it is seen that at finite temperature the change in energy due to finite size effects is not significant.
%(system sizes are increased by $3$ due to constraints in the parallelization algorithm). 
\begin{table}[htb]
\centering
\begin{tabular}{ |c | c |c| }
    \hline
    L & $\langle \mathcal{E} \rangle (\mathcal{T}=0.2)$ & $\langle \mathcal{E} \rangle (\mathcal{T}=0.4)$  \\ \hline \hline
    
    39& -2.31898& -2.14408\\
    42& -2.32090& -2.14087\\
    45& -2.32448& -2.13814\\
    48& -2.31905& -2.13881\\  
  
    \hline
  \end{tabular}
\caption{Ensemble energies for  $\langle \mathcal{E} \rangle$ at $ \mathcal{T}=0.2$ and  $\mathcal{T}=0.4$.  }
\label{SizeEng}
\end{table}
\\
 $\langle | \vec{S}_k | \rangle$ was also calculated and the results are shown in figure \ref{SizeCom}. If the spacing between Skyrmions was to change with temperature, one expects a small change in system size could stabilize six fold ordering at higher $T$. Systems with size $L$ equal to 45 or 48 show the same general trend in which the six order is destroyed leaving peaks in the $[1,1]$ and $[-1,1]$ directions. For $L=39$ the system doesn't form six fold packing. These results are consistent with the above interpretation of the phase transition as a result of changing Skyrmion profile.
\section{Conclusions and Comments}
We have shown that since the chiral density changes smoothly over a large temperature range, it does not capture all of the information about the melting process from the $SC_h$ phase in chiral magnets. The Fourier intensity and the reversed connected components of the system reveal a low temperature reduction in the chiral charge due to thermal distortion of the Skyrmion profiles. At higher temperatures there is a sharp loss of six fold order associated with creation of elongated structures that disrupt ordering. Only the ends of these structures contribute to the chiral charge, further reducing the chiral density. In addition it has been shown that the nature of hexagonal close packing means that while any choice of simulation size will introduce finite size effects, judicious choice of system size can help alleviate these effects.
\\
Here the authors have limited the scope of investigation to a single choice of applied field,
% in order to elucidate the concept of connected components as a analysis technique. 
however, the techniques presented might offer insight for a wide variety of fields. Specifically there is the possibility that at higher fields the relative strength $\sum_{i} \vec{H}.\vec{s}_i$ might suppress the creation of extended regions of reversed magnetization. There exists also the possibility of analyzing the zero field ground state as stripes of alternating connected regions of perpendicular spins pointing in the positive and negative $z$ direction,  represented as $\Theta(1/2( \pm \vec{s}_{i}.\hat{z} +1) -T_r)$. Similarly the $SC_1$ and $SC_2$ phase might be analyzed as a checkerboard alternating cores.
\\
 Interestingly the system maintains a net chiral charge at temperatures at which any long range ordering is destroyed which could have consequences for the calculation of hall conductivity at these temperatures \cite{PhysRevB.80.054416}. 
\section*{Acknowledgments}
The authors acknowledge funding from the Australian government department of Innovation, Industry, Science and Research, the Australian Research Council, the University of Western Australia and the Scottish Universities Physics Alliance.
%\section*{References}
%\begin{thebibliography}{99}
\bibliographystyle{unsrt}	% (uses file "plain.bst")
\bibliography{latex8}
%-\end{thebibliography}
\end{document}